# Tunable Inter-Moiré Physics in Consecutively-Twisted Trilayer Graphene

Wei Ren[1], Konstantin Davydov[1], Ziyan Zhu[2], Jaden Ma[3], Kenji Watanabe[4], Takashi Taniguchi[5], Efthimios Kaxiras[6,7], Mitchell Luskin[8], Ke Wang[1*]

[1]*School of Physics and Astronomy, University of Minnesota, Minneapolis, Minnesota 55455, USA*

[2]*Stanford Institute for Materials and Energy Sciences, SLAC National Accelerator Laboratory, Menlo Park, CA 94025, USA*

[3]*Department of Physics, The Ohio State University, Columbus, Ohio 43210, USA*

[4]*Research Center for Electronic and Optical Materials, National Institute for Materials Science, 1-1 Namiki, Tsukuba 305-0044, Japan*

[5]*Research Center for Materials Nanoarchitectonics, National Institute for Materials Science, 1-1 Namiki, Tsukuba 305-0044, Japan*

[6]*Department of Physics, Harvard University, Cambridge, Massachusetts 02138, USA*

[7]*John A. Paulson School of Engineering and Applied Sciences, Harvard University, Cambridge, Massachusetts 02138*

[8]*School of Mathematics, University of Minnesota, Minneapolis, Minnesota 55455, USA*

**We fabricate a twisted trilayer graphene device with consecutive twist angles of 1.33º and 1.64 º, in which we electrostatically tune the electronic states from each of the two co-existing moiré superlattices and the interactions between them. When both moiré superlattices contribute equally to the electrical transport, we report a new type of inter-moiré Hofstadter butterfly. Its Brown-Zak oscillation corresponds to one of the intermediate quasicrystal length scales of the reconstructed moiré of moiré (MoM) superlattice, shedding new light on emergent physics from competing atomic orders.**

When two layers of a vdW material are placed on top of each other misaligned by a small twist angle, the band structure of the beating pattern (moiré superlattice) can host a high density of states (DOS) arising from near dispersionless (flat) bands. In an interacting electron context, this diverging DOS drives the system toward behavior where electron correlations and localization become dominant, leading to exotic emergent quantum phenomena distinctly different from the behavior of the original material, including superconductivity, ferromagnetism, correlated insulator and quantum anomalous Hall states [1–10]. In twisted trilayer graphene (tTLG), where a third graphene layer is added on top of a conventional twisted bilayer graphene (tBLG) [11–18], a second twist angle becomes available for additional combinations [19]. The richness of possibilities with two independent twist angles elevates tTLG to a system of uniquely different material properties: (1) when the two twist angles *alternate,* that is, are equal in magnitude but opposite in direction (twist back), the original tBLG moiré states are preserved but enhanced with higher critical temperature and electrostatic tunability [11,12,15]; (2) when the two twist angles are *consecutive* (continued twisting), the transport signature of new families of quantum states was discovered at an extremely low carrier density on the order of $10^{10}$ cm$^{-2}$, arising from a higher order MoM superlattice [13]. While case (1) is well understood in the context of moiré physics, the microscopic mechanisms for the behavior in case (2) remain elusive due to the intrinsic complexity of the underlying atomic structure. The two sets of moiré superlattices that underlie case (2) behavior interact to give rise to a plethora of emergent length scales [13,20–22], and both contribute and compete in determining the transport behavior. Similar atomic reconstructions have also been reported in hBN/graphene/hBN sandwiches, where a single piece of graphene is modulated by hBN, instead of more complex competition between three equally deformable graphene layers [21-23].

Here we systematically study the interplay between the two moiré superlattices and their mutual role in determining the emergent quantum phenomena in a dual-gated consecutively-twisted tTLG device. Fig. 1a shows the schematic representation of the tTLG stack, where three pieces of monolayer graphene flakes are sequentially picked up and transferred on top of each other, with two consecutive twist angles of 1.33°±0.03° (1.64°±0.04°) between the top and middle (middle and bottom) layers. The two twist angles are chosen to be different so that the transport signature of each moiré superlattice can be isolated, corresponding to (see Fig. 1b schematic) a moiré superlattice constant of 8.6 nm for the top moiré (yellow) superlattice and 10.6 nm for the bottom moiré (purple) superlattice. Both twist angles are sufficiently different from the previously-reported tBLG magic angle, and correlated insulator and superconducting states are not expected to emerge from any of the individual moiré superlattice.

The sample is etched into the Hall bar geometry for magneto-transport measurements (Fig. 1b), conducted in a $^3$He/$^4$He dilution refrigerator at a base temperature of ~ 10 mK. A global Si bottom gate ($V_b$) and a graphite top gate ($V_t$) are used for tuning the overall carrier density $n$ in tTLG via capacitive gating with total gate potential ($V_b + \alpha V_t$, where $\alpha$ is the capacitive ratio between bottom and top gate). In addition, the two gates are used for tuning the relative distribution of such charge carriers among the two moiré superlattices ($V_b - \alpha V_t$) via an out-of-plane electrical field (displacement field $D$, positive when $V_b > \alpha V_t$). The dual-gate geometry and the addition of an out-of-plane electric field allows us to control the carrier distribution among the three layers. This enables us to enhance or suppress the contribution from each individual moiré superlattices to the transport behavior.

Fig. 1d shows the measured 4-probe longitudinal ($R_{xx}$) and transverse resistance ($R_{xy}$) as a function of out-of-plane magnetic field $B$, and moiré filling factors $v_1 = 4n/n_{10}$ ($v_2 = 4n/n_{20}$). Here $n_{10}$ ($n_{20}$) corresponds to carrier density when 4 charge carriers occupy per unit cell of the top (bottom) moiré superlattices, denoted as $v_1 = \pm 4$ ($v_2 = \pm 4$), where satellite fans emerging from band-insulator states of the top (bottom) moiré superlattice are observed. At $v_1 = \pm 4$ ($v_2 = \pm 4$), the transport features of a Landau fan diagram are noticeably less well-defined compared to that expected from an individual tBLG moiré superlattice [1–3], due to the presence of the additional bottom (top) graphene layer disrupting the Landau-level formation in the top (bottom) moiré superlattice. The band insulator states ($v = \pm 4$) of each moiré superlattice also demonstrates semi-metallic temperature dependence (see SI section S4), due to the third graphene layer and the absence of a gap at the edge of each set of moiré bands. The satellite Landau fan indices follow a sequence of tTLG (2, 6, 10,...) [11,12], in contrast to that from an individual tBLG moiré superlattice (4, 8, 12...) [1–3], confirming the crucial role of the extra graphene layer in shaping the transport signature of each moiré superlattice.

The behavior of the tTLG system has its unique features, different from those of tBLG, see Fig. 1c where $K_{L1}$, $K_{L2}$ and $K_{L3}$ denote the $K$ points of the top, middle and bottom graphene layers. Bands belonging to top and bottom moiré superlattices hybridize and compete in determining the overall transport behavior of the tTLG. The red (blue) color of the bands marks the polarization (relative concentration) of charge carriers in the top (bottom) moiré superlattices, with red (blue) color corresponding to carriers residing exclusively in the top (bottom) and middle graphene layers. At band insulator states of the top (bottom) moiré, denoted by yellow (purple) dashed line, a satellite fan is observed, with its transport features compromised by the interference from the bottom (top) moiré (whose band is partially filled).

To confirm the above picture, we demonstrate that this interference is tunable by a displacement field, $D$. For ease of demonstration, we first tune the device into the regime where the top moiré state dominates the transport over the bottom moiré state (in other words, charges are primarily occupying the top and middle graphene layers) and measure the satellite fan from the top moiré state, while tuning the relative contribution of the bottom moiré state with the $D$ field.

Starting with the electron-side of the top moiré fan near $v_1 = 4$ at $D = -0.11$ V/nm, the Shubnikov-de Hass (SdH) oscillation from the first Landau level appears at around $B = 2$ T (see Fig. 2e). When the displacement field becomes more negative, at $D = -0.22$ V/nm, electrons move towards the top moiré superlattice (Fig. 2f). The transport signature from the Landau levels in the top moiré superlattice becomes more prominent with reduced interference from the bottom layer. As a result, the SdH oscillation emerges at a lower magnetic field of $B = 0.5$ T. When the displacement field becomes more positive at $D = 0$ V/nm, electrons move closer to the bottom moiré superlattice (Fig. 2d), enhancing its interference, and resulting in delayed emergence of the SdH oscillation at $B = 3$ T.

For the hole-side of the top moiré fan near $v_1 = -4$, the dependence on $D$ is the opposite (Fig. 2a-c), as expected from this physics picture. As $D$ increases (decreases), the hole type charge carrier migrates towards (away from) the top moiré superlattice, leading to a more pronounced (compromised) satellite fan diagram with the SdH oscillation starting at a lower (higher) B field.

To study emergent physics from the co-existing moiré superlattices, we next measured the transport features at the intermediate carrier density range between $v_1 = +4$ and $v_2 = +4$, where a transport signature from both satellite fans is expected. The relative contribution from the two moiré patterns can be fine-tuned by the $D$ field (see Fig. 3). When a negative $D$ field is applied (Fig. 3a, b), the electrons move towards the top moiré superlattice, whose fan becomes more prominent than the feature from the bottom moiré superlattice. In contrast, when a positive $D$ field is applied, the electrons move towards the bottom moiré superlattice, whose fan is now more visible at the cost of a smeared-out top moiré fan (Fig. 3d, e). At $D = 0$ V/nm (Fig. 3c), the device is tuned into the strong-coupling regime, where both moiré patterns contribute equally, and therefore compete in determining the transport signature.

Fig. 4a shows a zoom-in high resolution scan of the measured 4-probe conductance, where a complex emergent pattern is observed, which we refer to as a new type of "inter-moiré Hofstadter butterfly". Longitudinal conductance peaks are observed at $B$ = 7.4 T, 4.9 T, 3.7 T, 3.0 T, 2.5 T, 2.1 T, … shown as horizontal lines (dashed) in 4-probe resistance (Fig. 4a). The inset shows $\sigma_{xx}$ as a function of $B$, averaged over the entire carrier density range of the color plot, signifying the magnetic field at which local conductance maxima are found along horizontal lines in the color plot (marked by arrows in inset and dashed lines in color plots), or Brown-Zak oscillations [23,24]. These magnetic field values correspond to 1/n (where n = 2, 3, 4, 5, 6, 7…) flux quanta ($\phi_0$) for a unit cell size of $283 \pm 7$ nm$^2$ (or lattice constant of $18.1 \pm 0.2$ nm.). In addition, a high-order conductance peak is also observed at a magnetic field corresponding to 2 flux quanta per 5 of such unit cells [23].

The previously reported Hofstadter butterfly in hBN-graphene or tBLG [25–28] is observed when fans from the satellite peak and the Dirac peak overlap, whose periodicity in $1/B$ is a consequence of emergent moiré periodicity. In contrast, the inter-moiré Hofstadter butterfly results from two sets of satellite fans crossing over (without participation of the main fan from charge neutrality point), each belonging to a superlattice with distinct lattice constant. The inter-moiré butterfly is therefore a signature of higher order superlattice periodicity reconstructed from two co-existing moiré superlattices. The unit cell size of $283 \pm 7$ nm$^2$ extracted from the inter-moiré butterfly agrees with one of the intermediate length scales in the relaxed atomic landscape of tTLG (shaded rhombus in Fig. 4b, with lattice vectors labeled by red arrows), one that is most prominent among a plethora of length scales that co-exist in consecutive-twisted trilayer graphene (Fig. 4c) [19]. Notably, the MoM superlattice has a unit cell size (~1527 nm$^2$) that is ~5.4 times larger, with a corresponding Brown-Zak oscillation expected at magnetic field values of $B = 2.70$ T, 1.34 T, 0.90 T, 0.67 T, 0.54 T, 0.46 T…, before SdH oscillations of the two competing fan diagram start to emerge.

In conclusion, we measure magneto-transport in a consecutively twisted trilayer graphene device, with twist angles of 1.33º±0.03º and 1.64º±0.04º. We observe two sets of satellite Landau fans, each belonging to one of two coexisting moiré superlattices. The two moiré superlattices compete in determining the transport signature of the tTLG at carrier density on the order of $10^{12}$ cm$^{-2}$, or typical moiré filling. We show that the strength and the hierarchy of the inter-moiré interaction can be tuned electrostatically, confirming the underlying physics picture. When the transport signatures of two moiré superlattices are equally present, we observe a new type of inter-moiré Hofstadter butterfly, the periodical pattern of which agrees with a higher order quasi-lattice reconstructed from the two moiré superlattices. Our work provides a comprehensive understanding of the complex interaction between coexisting atomic orders in tTLG, and how higher order periodicities and their transport signatures emerge from such competition. This opens a door towards understanding the intriguing microscopic mechanisms of emergent quantum phenomena in twisted multilayer 2D material platforms.

We thank Andrey Chubukov, Boris Shklovski and Yi Huang for helpful discussion. This work was supported by NSF DMREF Award DMR-1922165 and Award DMR-1922172, ARO MURI Grant No. W911NF-14-1-0247 and Simons Foundation Award No. 896626. Nanofabrication was conducted in the Minnesota Nano Center, which is supported by the National Science Foundation through the National Nano Coordinated Infrastructure Network, Award Number NNCI -1542202. Z.Z. is supported by a Stanford Science fellowship. Portions of the hexagonal boron nitride material used in this work were provided by K.W and T. T. K.W. and T.T. acknowledge support from the JSPS KAKENHI (Grant Numbers 20H00354, 21H05233 and 23H02052) and World Premier International Research Center Initiative (WPI), MEXT, Japan

[*]kewang@umn.edu

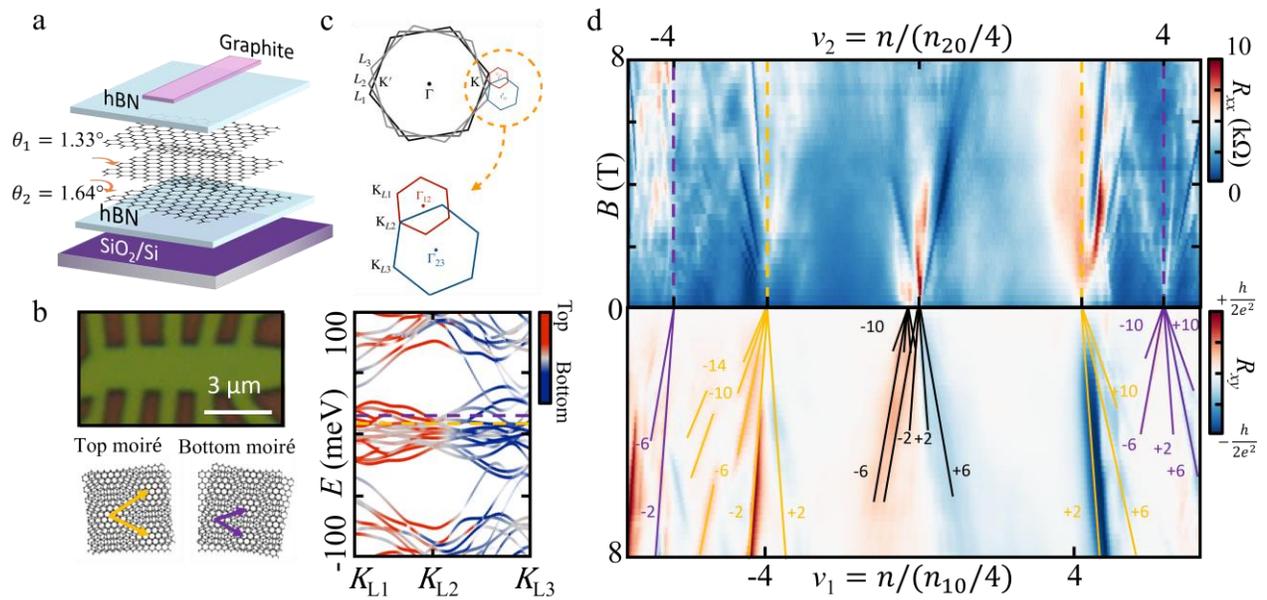

**Figure 1. Co-existing Moiré Superlattices in tTLG.** (a) Schematic image of the graphite-gated tTLG stack. (b) Upper: optical image of the Hall bar sample. Lower: schematic images of the top (bottom) moiré superlattice, whose lattice vectors are drawn as yellow (purple) arrows. (c) Upper: schematic images of the Brillouin zone of the three individual MLG layers. The zoom-in image shows the two moiré mini-Brillouin zones with different sizes formed between the adjacent bilayer pairs. Lower: Calculated band structure. The red and blue colors represent the layer polarization of charge carriers. The energy levels corresponding to $v_1 = \pm 4$ ($v_2 = \pm 4$) are indicated by the yellow (purple) dashed lines. (d) The measured 4-probe resistance $R_{xx}$ and $R_{xy}$ as a function of top (bottom) moiré filling factors (number of charges per moiré unit cell) and the magnetic field B. Two sets of satellite fans are observed to originate from the band-insulator states of the top (bottom) moiré superlattice, indicated by yellow (purple) lines.

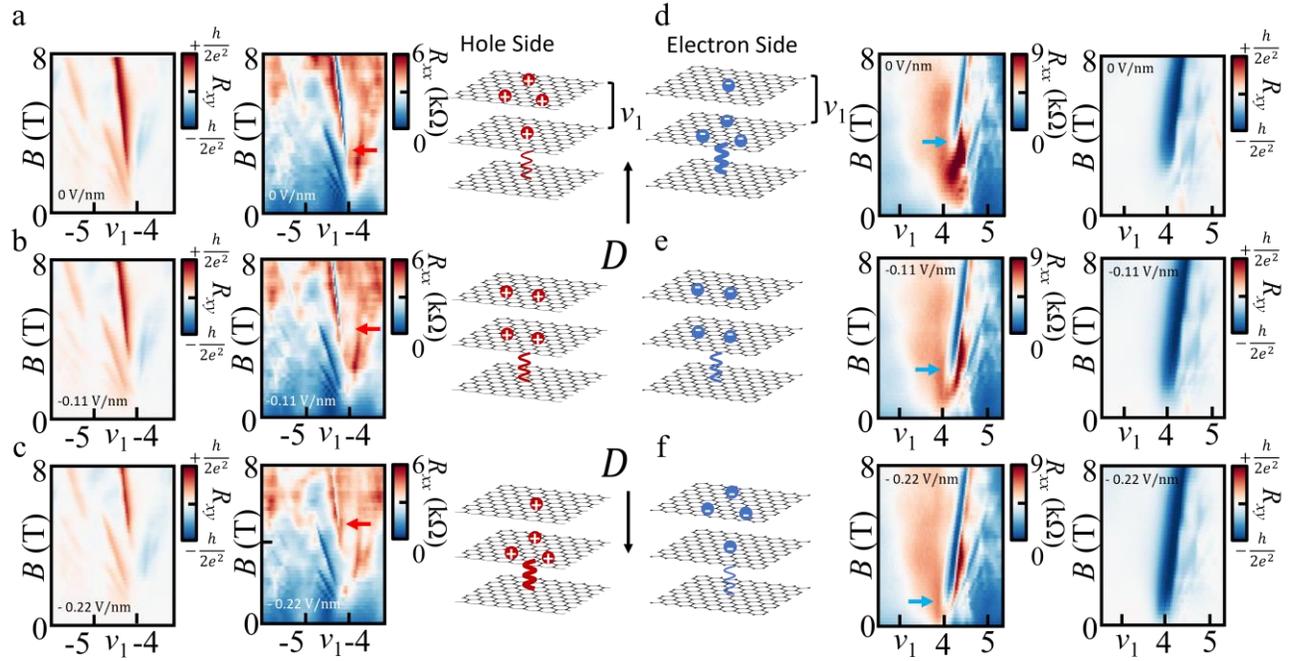

**Figure 2. Tunable Proximity Effect between Two Moiré Superlattices** (a-c) Measured 4-probe transverse resistance $R_{xy}$ (left), longitudinal resistance $R_{xx}$ (middle) near the hole-type band-insulator state of the top moiré at $v_1 = -4$, and schematic images of relative charge distribution (right) at (a) $D = 0$ V/nm, (b) $D = -0.11$ V/nm, and (c) $D = -0.22$ V/nm. The magnetic field at which the SdH oscillation starts to emerge decreases with $D$, as charges in top moiré superlattice moves away from the interference of the bottom graphene. (d-f) schematics of charge distribution (left), $R_{xx}$ (middle), and $R_{xy}$ (right) near the electron-type band-insulator state at $v_1 = +4$ at (d) $D = 0$ V/nm, (e) $D = -0.11$ V/nm, and (f) $D = -0.22$ V/nm. In contrast to the hole side, the transport signature of the satellite fan becomes less pronounced as $D$ increases, as electrons in the top moiré superlattice towards the bottom graphene and enhanced the hybridization.

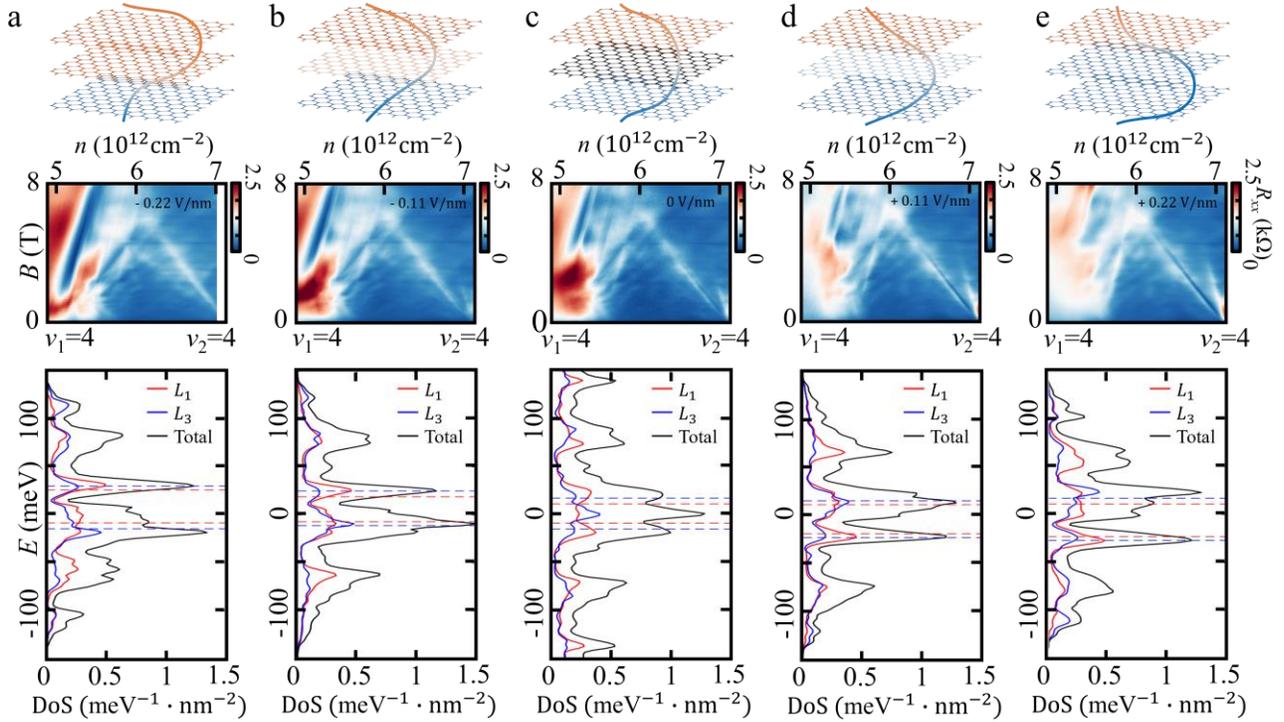

**Figure 3. Tunable Hierarchy between two Moiré Superlattices.** The top (bottom) graphene competes for the middle layer in dominating the transport signature with that from top (bottom) moiré superlattice. The result of such competition can be tuned by displacement field $D$ = (a) $-0.22$ V/nm, (b) $-0.11$ V/nm, (c) $0$ V/nm, (d) $+0.11$ V/nm, (e) $+0.22$ V/nm. (Top row) The electrons in the middle graphene move closer to the top (bottom) graphene as $D$ increases (decreases), participating more in pronouncing the transport signature of the top (bottom) moiré, with the color denoting the graphene layer is inclined towards. (Curve) Cartoon of electron wavefunction distribution along out-of-plane axis. (Middle row) We demonstrate the 4-probe resistance $R_{xx}$, measured at carrier density in between $v_1 = +4$ and $v_1 = +4$, where the transport signature from both moiré superlattices is expected. As $D$ increases, the satellite fan of the winning top (bottom) moiré lattice is more visible, at the cost of the bottom (top) moiré fan being suppressed. This is consistent with the (bottom row) calculated density of states in the top and bottom moiré superlattice, where the red, blue, and black curves corresponding to the density of states in the top, bottom graphene layer and in the entire tTLG. The red (blue) dashed lines indicate the energies corresponding to $v_1(v_2)= \pm 4$.

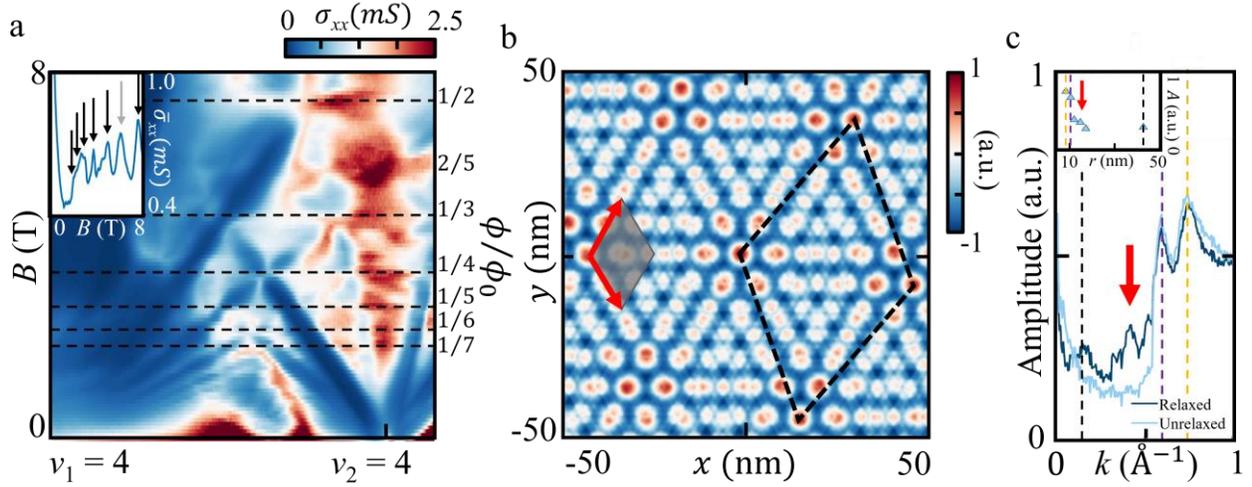

**Figure 4. Inter-moiré Hofstadter Butterfly.** (a) Conductance $\sigma_{xx}$ as a function of filling factors and the magnetic field (left axis) or magnetic flux $\phi = BS$ in the unit of flux quantum $\phi_0$ (right axis). Sequence of $\phi/\phi_0 = 1/n$ is outlined by black dashed lines. Insert: $\sigma_{xx}$ as a function of $B$, averaged over the entire carrier density range of the color plot, signifying the magnetic field at which conductance peaks are found along horizontal lines in the color plot (marked by arrows in inset and dashed lines in color plots). (b) Simulation of relaxed atomic landscape of tTLG, with the color scale plotting relaxed total misfit energy landscape. The periodicity of the inter-moiré Brown-Zak oscillation (in $1/B$) corresponds to a unit cell (grey shade) size $283 \pm 7$ nm$^2$, or lattice (red arrows) constant of $18.1 \pm 0.2$ nm. (c) The length scale agrees with one of the most prominent among a plethora of co-existing length scales in tTLG, shown by calculated Fourier spectrum of relaxed tTLG. Peaks are found at dominant length scales at lattice constant of MoM superlattice (black dashed, 42.8 nm), top moiré superlattice (orange dashed, 10.5 nm), bottom moiré superlattice (blue dashed, 8.4 nm), and additionally at intermediate length scale between 15 - 20 nm (arrow) responsible for the observed Brown-Zak oscillation. Inset: Dominate length scales and their relative weight.

# Tunable Inter-Moiré Physics in Consecutively-Twisted Trilayer Graphene

Wei Ren[1], Konstantin Davydov[1], Ziyan Zhu[2], Jaden Ma[3], Kenji Watanabe[4], Takashi Taniguchi[5], Efthimios Kaxiras[6,7], Mitchell Luskin[8], Ke Wang[1*]

[1]*School of Physics and Astronomy, University of Minnesota, Minneapolis, Minnesota 55455, USA*

[2]*Stanford Institute for Materials and Energy Sciences, SLAC National Accelerator Laboratory, Menlo Park, CA 94025, USA*

[3]*Department of Physics, The Ohio State University, Columbus, Ohio 43210, USA*

[4]*Research Electronic and Optical Materials, National Institute for Materials Science, 1-1 Namiki, Tsukuba 305-0044, Japan*

[5]*Research Center for Materials Nanoarchitectonics, National Institute for Materials Science, 1-1 Namiki, Tsukuba 305-0044, Japan*

[6]*Department of Physics, Harvard University, Cambridge, Massachusetts 02138, USA*

[7]*John A. Paulson School of Engineering and Applied Sciences, Harvard University, Cambridge, Massachusetts 02138, USA*

[8]*School of Mathematics, University of Minnesota, Minneapolis, Minnesota 55455, USA*

## S1. Sample Preparation and Device Fabrication

The dual-gated twisted trilayer graphene (tTLG) devices are made by the 'cut and tear' method [1]. hBN, graphite and monolayer graphene flakes are exfoliated [2] and characterized by the atomic force microscope (AFM) to be atomically clean. A single piece of monolayer graphene was cut into three individual pieces with the same lattice orientation by the atomic force microscope (AFM). With the help of a poly (bisphenol A carbonate) (PC) and polydimethylsiloxane (PDMS) stamp on a glass slide [3], we pick up the first piece of hBN and a few-layer graphite flake (serving as a top gate) and the second piece of hBN (serving as dielectrics between the graphite gate and underline tTLG). Then the three pieces of the precut graphene are picked up consecutively, each time with a twist angle in the same direction. The third piece of hBN is picked up to encapsulate the tTLG and the whole stack is transferred onto a clean $SiO_2$(285 nm)/Si substrate (serving as a back gate) at 180℃ [4]. After the PC residue on the top hBN surface is cleaned by chloroform, acetone and isopropanol, the Cr/Pd/Au (1 nm/5 nm/180 nm) metal contacts are added to the sample by e-beam lithography, plasma etching and e-beam evaporation processes [5]. Finally, a Hall-bar shaped bubble-free region of ~ 3 μm × 8 μm (Sample #1) (~ 1 μm × 6 μm (Sample #2)) is defined by e-beam lithography and plasma etching.

Following the above method, we fabricate two consecutively twisted trilayer graphene (tTLG) samples: Sample #1 with $\theta_1 = 1.33°\pm0.03°$ and $\theta_2 = 1.64°\pm0.04°$ ($\theta_1/\theta_2 \approx 4/5$), and Sample #2 (without the AFM pre-cutting the MLG flake) with $\theta_1 = 1.88°\pm0.01°$ and $\theta_2 = 1.24°\pm0.01°$ ($\theta_1/\theta_2 \approx 2/3$). Sample

#1 has multiple device regions, which we will label as D1, D2, D3 and D4. S2 provides a summary of additional transport data from different regions in Sample #1 and S3 provides a summary of Sample #2.

## S2. Transport Signature in Other Device Regions of Sample #1

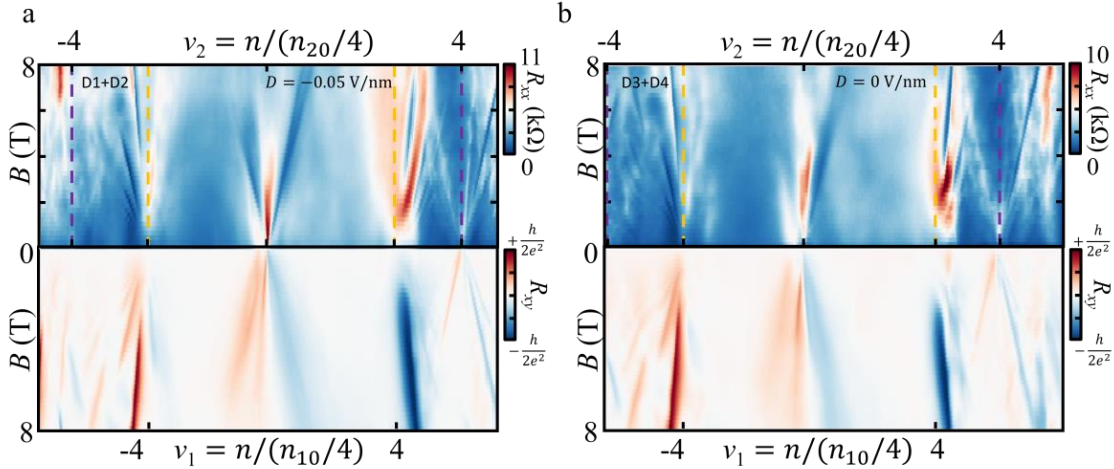

**Figure S1. Transport Signatures of Co-existing Moiré Superlattices in Other Device Regions of Sample #1.** The measured 4-probe resistance $R_{xx}$ and $R_{xy}$ as a function of top (bottom) moiré filling factors (number of charges per moiré unit cell) and the magnetic field B in (a) device region D1+D2 and (b) device region D2+D3. Two sets of satellite fans are observed to originate from the band-insulator states of the top (bottom) moiré superlattices, indicated by yellow (purple) lines.

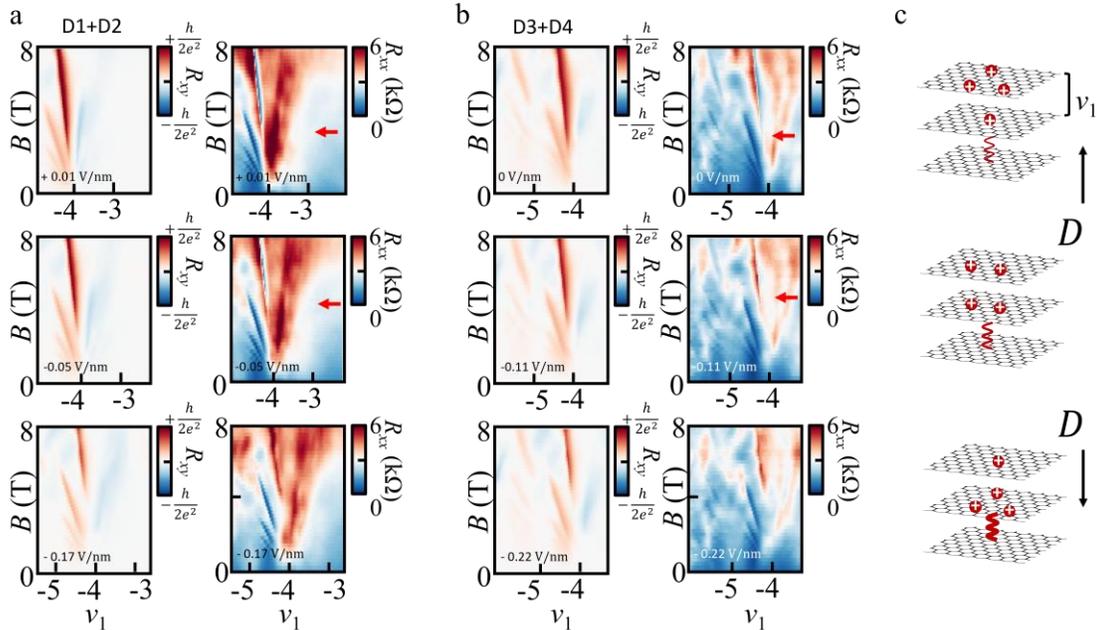

**Figure S2. Tunable Proximity Effect between Two Moiré Superlattices (Hole Side) in Other Device Regions of Sample #1.** Measured 4-probe transverse resistance $R_{xy}$ and longitudinal resistance $R_{xx}$ near the hole-type band-insulator state of the top moiré at $v_1 = -4$ in (a) Device Region D1+D2 and (b) Device Region D3+D4. (c) Schematic images of relative charge distribution (right) at various $D$ fields. The magnetic field at which SdH oscillation starts to emerge decreases with $D$, as charges in top moiré superlattice moves away from the interference of the bottom graphene.

Here we summarize additional data measured from other device regions which qualitatively reproduce the core physics picture. In the main manuscript, Fig. 1c and Fig. 2 are the transport data from a combined region D2 and D3 (3 μm × 4 μm ) in Sample #1 at zero perpendicular displacement field $D$. Fig. S1a (S1b) are the full range scans of longitudinal resistance $R_{xx}$ and transvers resistance $R_{xy}$ from the combined region D1 and D2 (D3 and D4), which we label as D1+D2 (D3+D4). Two sets of satellite fans from two distinct moiré superlattices are observable and indicated by yellow and purple dash lines, respectively. Fig. 2 shows the transport data from the combined region D2 and D3, demonstrating the electric field tunable proximity effect between two moiré superlattices. Fig. S2 (hole side) and Fig. S3 (electron side) provide additional data on device regions D1+D2 and D3+D4, reproducing the tunable proximity effect. To clearly see the co-existence of transport signature from both moiré superlattices, we choose to measure over smaller device regions (3 μm × 2 μm ) to avoid the effect brought by the angle inhomogeneity [6,7]. Fig. 3 is transport signature from device region D2 at intermediate carrier density in between $v_1 = +4$ and $v_2 = +4$, where transport signature from both moiré superlattices is expected. Fig. S4 provides a summary of transport signatures from device regions D1 (Fig. S4a), D3 (Fig. S4b), and D4 (Fig. S4c), where qualitatively reproduce the tunable hierarchy between two moiré superlattices. The effect from inhomogeneity across different device regions is also observable. D1 has the highest sample quality among all the regions and resolves satellite fans from both moiré superlattices clearly. Inter-moiré butterflies in other device regions (other than D1 as shown in the main manuscript) are summarized in Fig. S5.

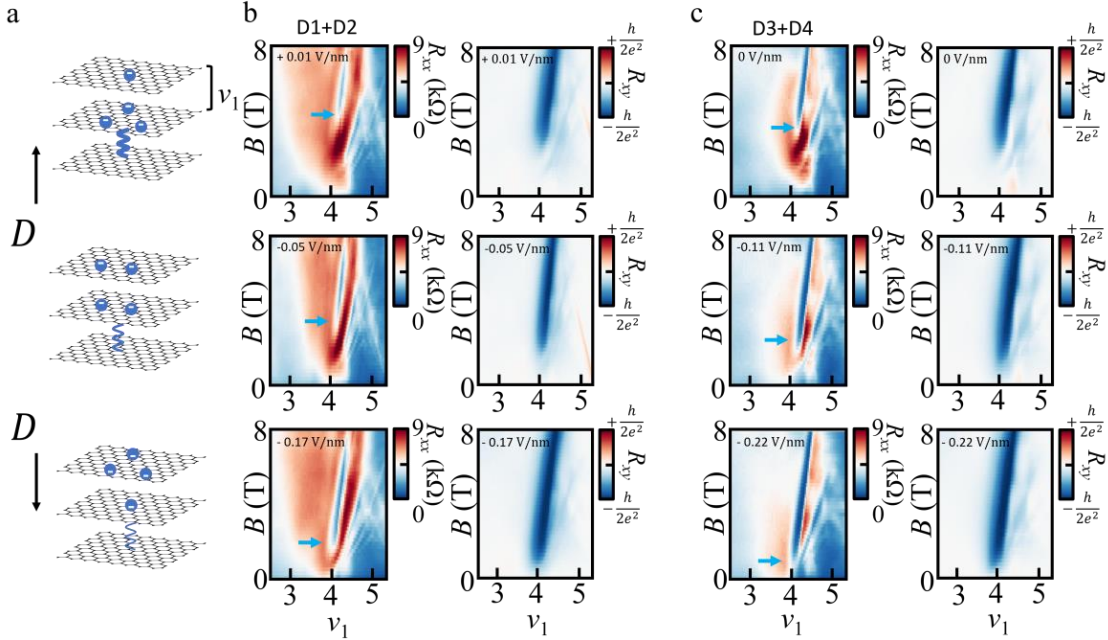

**Figure S3. Tunable Proximity Effect between Two Moiré Superlattices (Electron Side) in Other Device Regions in Sample #1.** (a) Schematic images of relative charge distribution (right) at various $D$ fields. Measured 4-probe transverse resistance $R_{xy}$ and longitudinal resistance $R_{xx}$ near the electron-type band-insulator state of the top moiré superlattice at $v_1 = +4$ in (b) Device Region D1+D2 and (c) Device Region D3+D4. In contrast to the hole side, the transport signature of the satellite fan becomes less pronounced as $D$ increases, as electrons in the top moiré superlattice towards the bottom graphene and enhanced the hybridization.

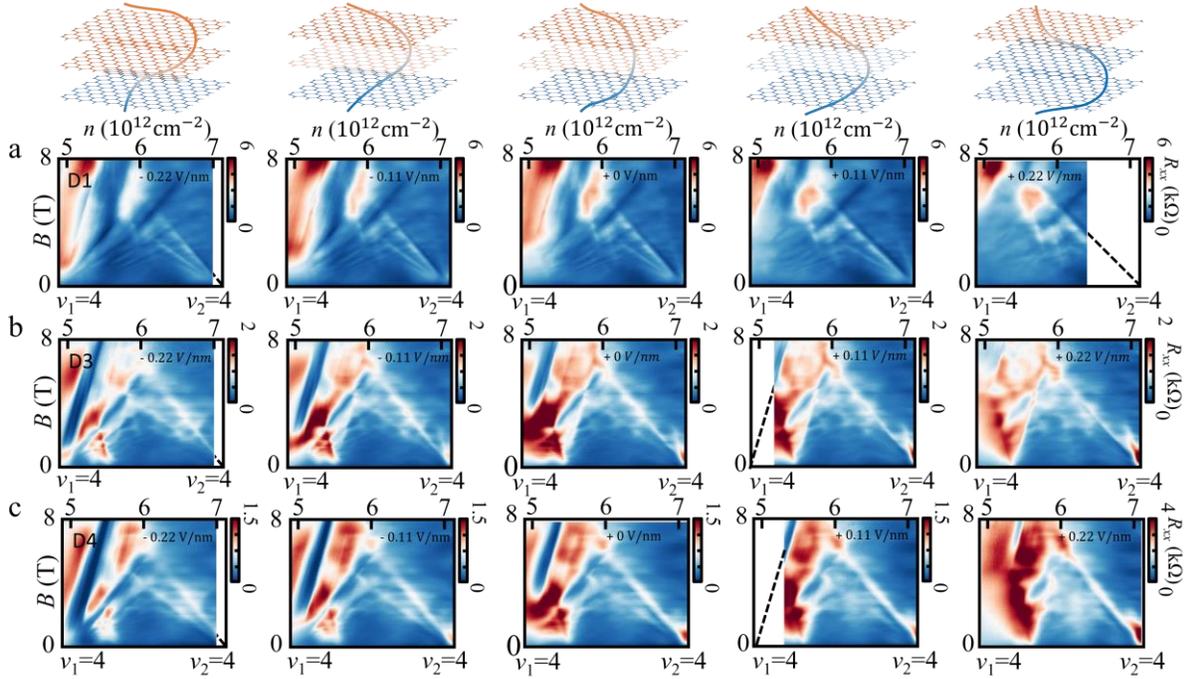

**Figure S4. Tunable Hierarchy between two Moiré Superlattices in Other Device Regions in Sample #1.** The measured 4-probe resistance $R_{xx}$ at carrier density in between $v_1 = +4$ and $v_2 = +4$ in Device Region (a) D1, (b) D3, and (c) D4, where transport signature from both moiré superlattices is expected. As $D$ increases, the satellite fan of the winning top (bottom) moiré superlattice is more visible, at the cost of the bottom (top) moiré fan being suppressed. This is consistent with the (bottom row) calculated density of states in the top and bottom moiré superlattice.

S

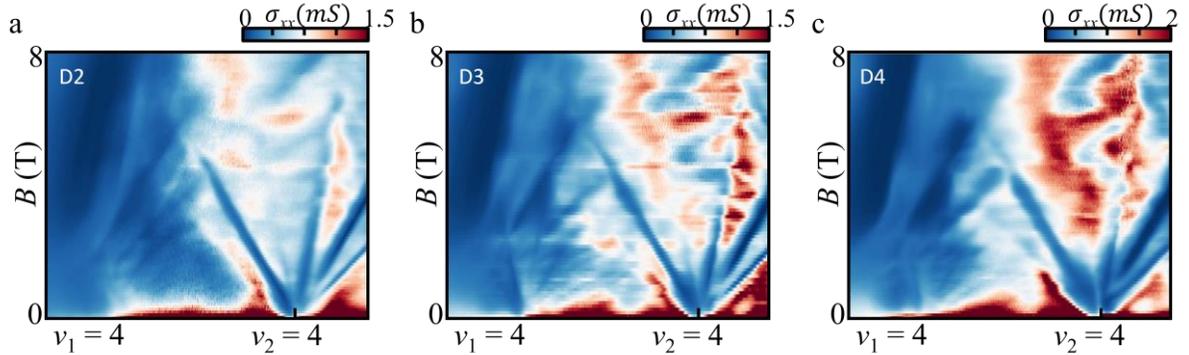

**Figure S5. Inter-moiré Hofstadter Butterfly in Other Device Regions in Sample #1.** Conductance $\sigma_{xx}$ as a function of filling factors and the magnetic field in Device Region (a) D2, (b) D3, and (c) D4, where the complicated inter-moiré butterfly is present.

### 3. Transport Data from Sample #2

The interplay between two moiré superlattices was also observed in a different tTLG sample, Sample #2 (Fig. S6), reproducing the qualitative behavior of Sample #1 from the main manuscript. The transport measurements of Sample #2 also showed two sets of satellite Landau fans (Fig. S6 c) signifying the presence of two moiré periodicities. However, in this case the top moiré superlattice has a smaller spatial periodicity than the bottom one, which is confirmed by the displacement field dependence of the device's resistance. The indices of the Landau levels at the top and bottom moiré band insulator states also follow the sequence (±2, ±6, ±10) typical for a tTLG system (Fig. S6c). By analyzing the position of the Landau fans, we characterize the twist angles of the top and bottom moiré superlattices to be 1.24° ±0.01° and 1.88°±0.01°. Similar to Sample #1, this system can be set in the regime where the proximity effect of a third layer on one of the moiré superlattices can be tuned by the displacement field (Fig. S7). By looking at the electron side of the Landau fan of the bottom moiré state ($v_2 = 4$) at $\Delta D = 0$ V/nm, (Fig. S7e). one can observe SdH oscillations emerging at $B = 1.5$ T. However, in contrary to the electrons at $v_1 = 4$ in Sample #1 (Fig. 1), the displacement field has the opposite effect. Applying a positive $\Delta D = 0.36$ V/nm displacement field (Fig. S7d) pushes the electrons into the bottom moiré inhibiting the proximity effect of the top layer. Thus, resulting in SdH oscillations becoming prominent at a lower magnetic field of 1T. In contrast, a negative displacement field $D = -0.36$ V/nm (Fig. S7f) brings the electrons in proximity to the very top layer. Hence, increasing its influence on the bottom moiré state and leading the SdH oscillations appear at an elevated ($B = 2.5$ T) magnetic field.

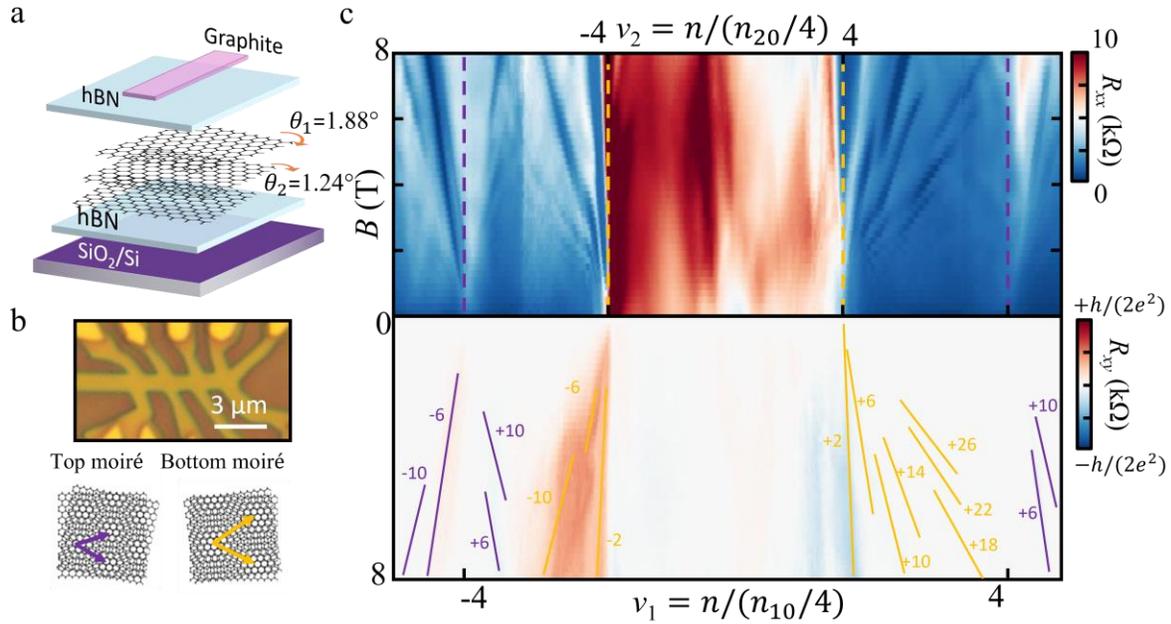

**Figure S6. Transport Signatures of Two Moiré Superlattices in Sample #2.** (a) Schematic view of the Sample #2 stack. (b) Upper: optical image of Sample #2. Lower : schematic images of the top (bottom) moiré periodicities. The lattice vectors of the two top (bottom) moirés are shown in purple (yellow). (c) The measured 4-probe $R_{xx}$ and $R_{xy}$ resistances as a function of top (bottom) moiré filling factors ($v_1$ and $v_2$ respectively) and the magnetic field. Two pairs of satellite fans are visible emanating from the band insulator states of the top (bottom) moirés marked by the purple (yellow) dashed lines. The band insulator nature of the fans is confirmed by the Landau level indices corresponding to a tTLG system.

The effect of the displacement field becomes reversed for the hole side of the ($v_2 = -4$) band insulator fan (Fig. S7a-c). The more positive (negative) grows the displacement field, the closer (further) the holes drift towards (from) the very top layer. Thus, increasing its interference with the bottom moiré.

The competition between the two moiré superlattices tunable by displacement field can also be observed in Sample #2. If one sets the system at the charge carrier density when features from both $v_1 = -4$ and $v_2 = -4$ fans coexist, a negative displacement field forces the holes to move into the bottom two layers (Fig. S8 a, b), yielding more SdH oscillations dips near the $v_2 = -4$ fan. In contrast, increasing the displacement field (Fig. S8 c-d), suppresses the contribution of the bottom moiré, smearing out its fan and leading to additional visible SdH oscillations from the top moiré fan.

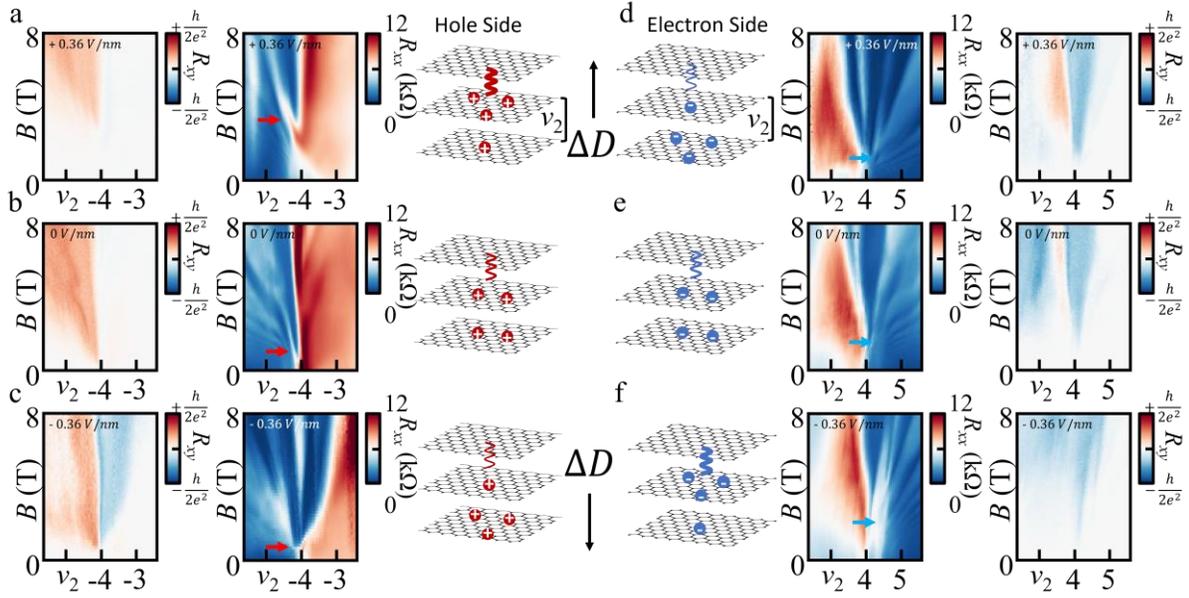

**Figure S7. Tunable Proximity Effect between Moiré Superlattices and Remaining Layer in Sample #2.** (a~c) Measured Hall $R_{xy}$ (left), and longitudinal $R_{xx}$ (middle) resistances under magnetic field near the $v_2 = -4$ hole side Landau fan at (a) $D = +0.36$ V/nm, (b) $D = 0$ V/nm, and (c) $D = -0.36$ V/nm. The right panels schematically show the hole redistribution between the layers under changing displacement field. The arrow marks the lowering magnetic fields of the SdH oscillations onset as the displacement field becomes more negative. (d~f) schematic images of negative charge carriers' distributions (left), $R_{xx}$ (middle), and $R_{xy}$ (right) of the electron side near $v_2 = +4$ at (d) $D = +0.36$ V/nm, (e) $D = 0$ V/nm, and (f) $D = -0.36$ V/nm. The blue arrow indicates the magnetic field at which the SdH oscillations become visible under varying displacement fields.

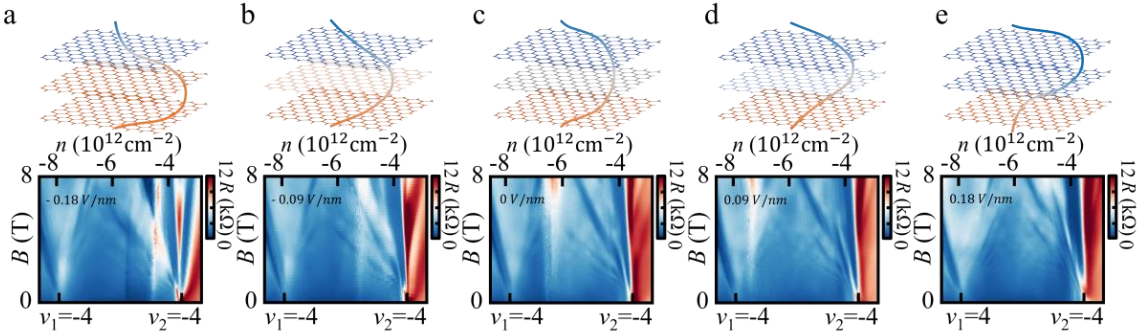

**Figure S8. Tunable Hierarchy between two Moiré Superlattices in Sample #2.** Upper: schematic images interlayer coupling in tTLG at different $D =$ (a) $-0.18$ V/nm, (b) $-0.09$ V/nm, (c) $0$ V/nm, (d) $+0.09$ V/nm, (e) $+0.18$ V/nm. The strength of coupling between two layers is indicated by the color: same color indicates the strong coupling regime between two adjacent layers. Lower: longitudinal resistance versus the charge carrier density and magnetic field at rising displacement field. The transport data shows the competition between the $v_1 = -4$ and $v_2 = -4$ Landau fans. Additional SdH features of the $v_1 = -4$ ($v_2 = -4$) fan become more prominent as the displacement field grows (decreases).

## S4. Detailed Transport Signature of Moiré Band-insulator States

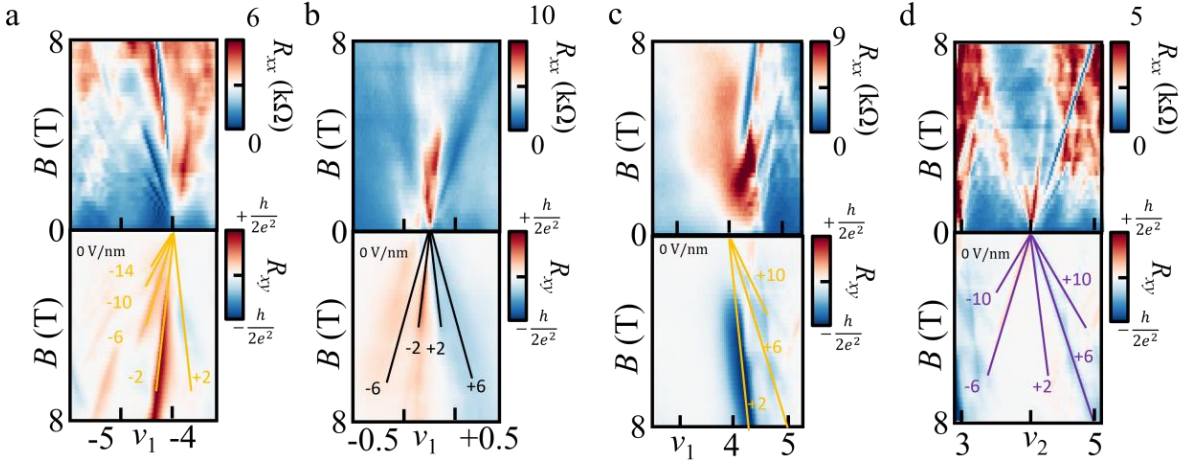

**Figure S9. Zoom in Scans of $R_{xx}$ and $R_{xy}$.** $R_{xx}$ (upper) and $R_{xy}$ (lower) of Sample #1 (device region D2+D3) as a function of moiré filling factors $v_{1,2}$ and magnetic field $B$ around (a) $v_1 = -4$, (b) $v_{1,2} = 0$, (c) $v_1 = +4$, and (d) $v_2 = +4$. Landau fan indices follow $\pm 2, \pm 6, \pm 10 ...$, which is typical of a tTLG or pure TLG sample.

In contrast to the Landau fan indices (4, 8, 12, 16…) at full moiré fillings (4 electrons or holes in each moiré unit supercell) from a standalone twisted bilayer graphene (tBLG) [3,8–11], the Landau fan indices in our consecutively twisted trilayer graphene samples are in series of 2, 6, 10, 14…. at $v_{1,2} = 0$, $\pm 4$ (Fig. S9 and Fig. 5c). This is consistent with previous reported trilayer graphene systems either with twisted [12–14] or without [15,16], signifying the crucial role of the extra graphene layer in shaping the transport signature from both top and bottom moiré superlattice.

Another effect brought by the extra graphene is the absence of a well-defined moiré band gap. The band structure calculation in Fig. 1d shows that there the density of states at $v_{1,2} = 0, \pm 4$ is finite instead of zero as in tBLG [17]. Intuitively, the third graphene layer shorts the current path when one of the moiré superlattices is insulating so that the entire sample behaves like a semi-metal as graphene. We can further confirm that there are no well-defined insulating states at charge neutrality point or $v_{1,2} = 0, \pm 4$ by examining the temperature dependence of the longitudinal resistance $R_{xx}$. Moiré full fulling states ($v_{1,2} = 0, \pm 4$) in Sample #1 (Fig. S10a) shows almost no temperature dependence from 10 mK to 20 K. The lack of exponential temperature dependence verifies that no insulating states at $v_{1,2} = 0, \pm 4$ are observed. Sample #2 yields similar results (Fig. S10b) except for $v_2 = -4$, where the $R_{xx}$ decreases with T but no strong exponential dependence on T is observed. $v_2 = -4$ is more of a semi-metallic state instead of insulating.

## S5. Definition of Zero Perpendicular Displacement Field

Due to residual charges and charge imbalance [18–20] between graphene layers in the hBN encapsulated tTLG samples, there can exist a finite perpendicular displacement field $D_0$ across the sample even without applying voltages on the graphite gate or Si backgate. In principle, a finite external $D$ field is needed to compensate for $D_0$, which requires additional efforts on characterizing $D_0$. Instead, per our need in the above physics picture, we defined the zero $D$ field in our paper as when transport signatures (satellite fans) from both moiré superlattices are most clear, where both moiré superlattices contribute equally to determining the transport signature. Intuitively, this is when charge carriers are equally distributed among the top and bottom graphene layer. Applying additional gate voltages therefore starts to redistribute the charge carriers away from this balanced configuration. Specifically, we choose Fig. 3c (Fig. S8c) as the zero $D$ field configuration for Sample #1 (#2). $D_0$ is equal to $0.11$ V/nm for Sample #1 and $0$ V/nm for Sample #2.

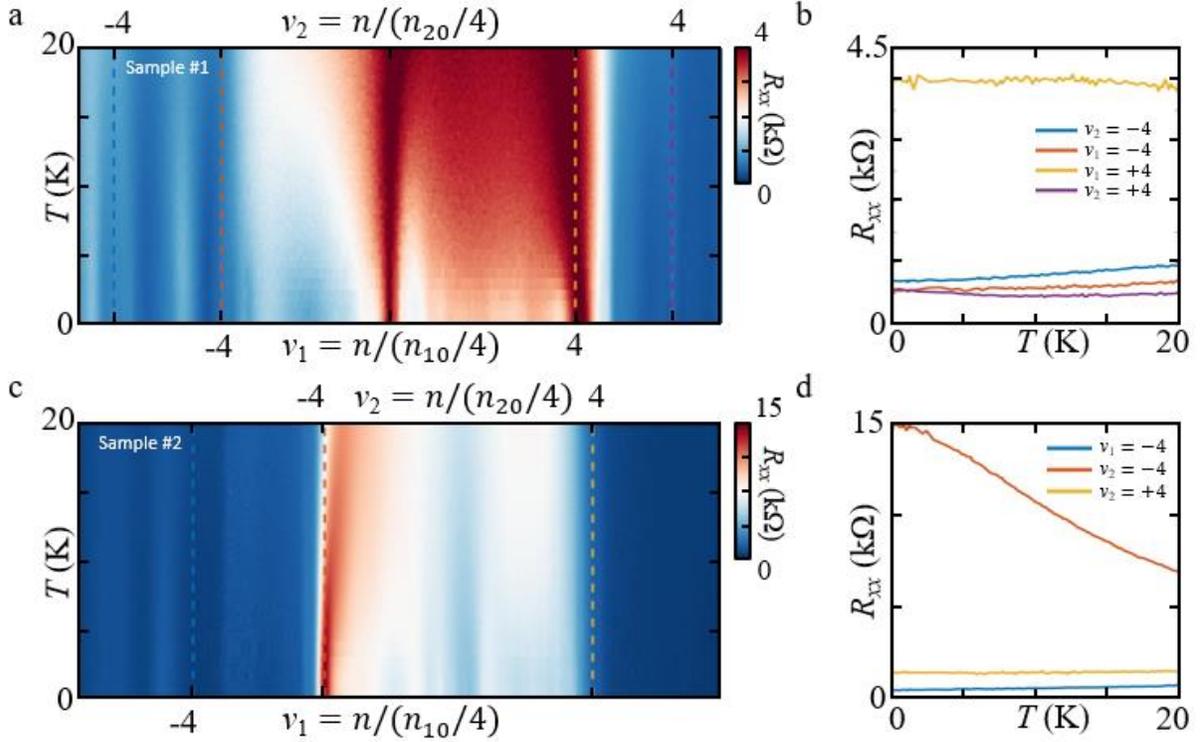

**Figure S10. Temperature Dependence of $R_{xx}$.** (a) $R_{xx}$ of Sample #1 (device region D1+D2) as a function of moiré filling factors $\nu_{1,2}$ and temperature $T$. (b) 1D cuts of (a) along $\nu_{1,2} = 0, \pm 4$, where almost no temperature dependence is observed. (c) $R_{xx}$ of Sample #2 as a function of moiré filling factors $\nu_{1,2}$ and temperature $T$. (d) 1D cuts of (a) along $\nu_{1,2} = 0, \pm 4$, where either no temperature dependence or no strong exponential temperature dependence is observed.

### S6. Evolution of Density of States in Each Moiré Superlattice under Displacement Fields

Following the method in section S7, the total density of states (DOS), the DOS in the top and bottom moiré superlattice at different displacement fields (Fig. S11) are calculated as supplementary information to the bottom panels in Fig. 3. The twisted angles used in the calculation are $\theta_1 = 1.33°$ and $\theta_2 = 1.64°$. Take the electron side as an example, the DOS in the top moiré superlattice increases as $D$ becomes more negative. This is consistent with the picture that electrons favor the top two layers and the experimental observation that the satellite fans emerging from $\nu_1 = +4$, which is the winning moiré superlattice, is more visible. Both top and bottom moiré is both electron and hole sides yield the same trend.

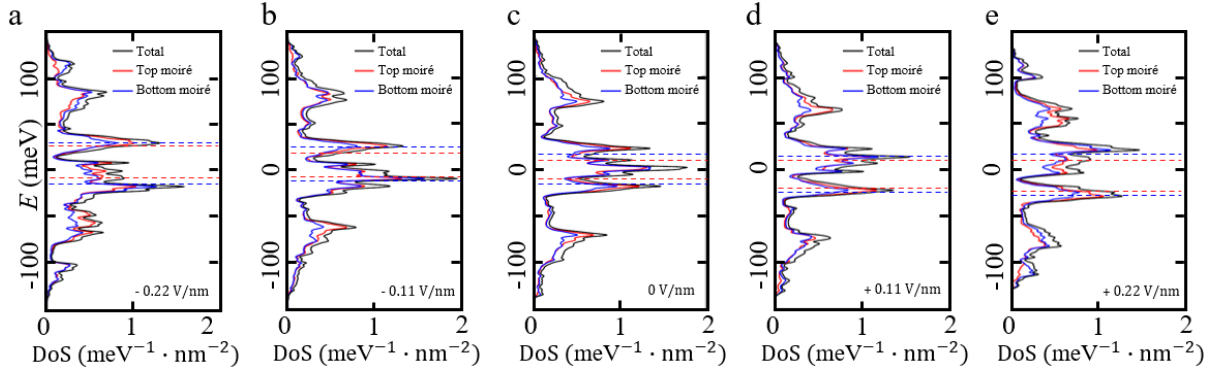

**Figure S11. Calculated Density of States in Each Moiré Superlattice at Different Displacement fields.** The total DoS (black curve), the DoS in the top moiré superlattice (red curve) and the bottom moiré superlattice (blue curve) under different displacement field $D$ = (a) $-0.22$ V/nm, (b) $-0.11$ V/nm, (c) 0 V/nm, (d) $+0.11$ V/nm, (e) $+0.22$ V/nm are calculated. The red (blue) dashed lines correspond to full moiré fillings $v_1 = \pm 4 (v_2 = \pm 4)$ where the satellite fans start to emerge. The evolution of DoS in each moiré is consistent with the fact that the satellite fan of the winning top (bottom) moiré superlattice is more visible as $D$ increases (Fig. 3).

## S7. Extracting Unit Cell Size from Inter-Moiré Butterfly

Conductance peaks are found at a series of $B$ field values in the Brown-Zak oscillations (Fig. 4a), where $BA/\phi_0 = 1/n$ is satisfied so they correspond to $1/n$ flux quanta ($\phi_0$) for a unit cell size of $A$. Then the peak position in terms of $1/B$ should satisfy $1/B = (A/\phi_0) \cdot n$. We plot the $1/B$ values of the leading peaks as a function of their corresponding index $n$, where $n$ is taken as inters from 2 to 7 and 5/2. The error bars on the $1/B$ values are estimated from the full-width at half-maximum of the conductance peaks. From the slope of a linear fit on the data, we extract $A$ to be $283 \pm 7$ nm$^2$.

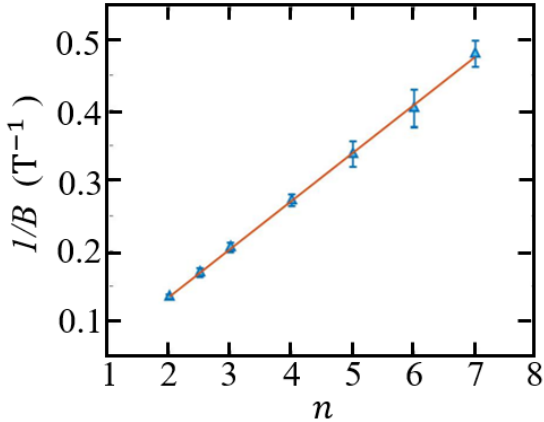

**Figure. S12 Subtracting Unit Cell Size with Linear fit.** Tracking the position of the conductance peaks, i.e., the corresponding $1/B$ value of the peaks, as a function of index $n$. $n$ is taken as inters from 2 to 7 and 5/2 in the plot. The error bars are esitmaed by the full-width at half-maximum of the peaks. The size of the unit cell that is corresponding to the Brown-Zak oscillation frequency can be extracted from the slope of the linear fit.

## S8. Band Structure Calculation

To calculate the band structure of twisted trilayer graphene, we adopt a momentum space continuum model developed in [21]. Here we briefly review the model. We start from a real-space tight-binding approximation for each individual layer. We consider the interlayer hopping between nearest-neighbor layers. We write the real-space tight binding Hamiltonian as

$$H = \sum_{\ell=1}^{3} H^\ell + \sum_{\ell=1,2} \left( V^{\ell,\ell+1} + V^{\ell,\ell-1} \right), \tag{S1}$$

where $H^\ell$ is the Hamiltonian for the $\ell$-th monolayer and $V^{\ell,\ell+1}$ describes the interlayer hopping between adjacent bilayer pairs. In a second quantized notation, $H^\ell$ can be written as

$$H^\ell = -t \sum_{R^{(\ell)}} c_{\ell,A}^\dagger \left[ c_{\ell,B}\left(R^{(\ell)}\right) + c_{\ell,B}\left(R^{(\ell)} - a_1^{(\ell)}\right) + c_{\ell,B}\left(R^{(\ell)} - a_2^{(\ell)}\right) \right], \tag{S2}$$

where $c_{\ell,\alpha}^\dagger$ and $c_{\ell,\alpha}$ are the creation and annihilation fermionic operator of the orbital $\alpha$ in layer $\ell$ and $a_i^{(\ell)}$ is the basis of the unit cell in the layer $\ell$. Similarly, the interlayer coupling term takes the form

$$V^{ij} = \sum_{R^{(i)},\alpha,R^{(j)},\beta} c_\alpha^\dagger(R^{(i)}) t_{\alpha\beta}^{ij}(R^{(i)}, R^{(j)}) c_\beta(R^{(j)}), \tag{S3}$$

where $t^{ij}(R^{(i)}, R^{(j)})$ is the overlap matrix element defined as

$$t_{\alpha\beta}^{ij}(R^{(i)}, R^{(j)}) = \langle i, R^{(i)}, \alpha | H_{ij} | j, R^{(j)}, \beta \rangle. \tag{S4}$$

Define the Fourier transform,

$$c_{\ell,\alpha}^\dagger(R^{(\ell)}) = \frac{1}{\sqrt{A_{\text{BZ},\ell}}} \int_{\text{BZ},\ell} d k^{(\ell)} e^{i k^{(\ell)} \cdot \left( R^{(\ell)} + \tau_\alpha^{(\ell)} \right)} c_{\ell,k^{(\ell)},\alpha}^\dagger, \tag{S5}$$

$$c_{\ell,\alpha}(R^{(\ell)}) = \frac{1}{\sqrt{A_{\text{BZ},\ell}}} \int_{\text{BZ},\ell} d k^{(\ell)} e^{-i k^{(\ell)} \cdot \left( R^{(\ell)} + \tau_\alpha^{(\ell)} \right)} c_{\ell,k^{(\ell)},\alpha}, \tag{S6}$$

where the integral is over the Brillouin zone of the $\ell$-th layer, and $\tau_\alpha^{(\ell)}$ describes the nearest neighbor separation between A and B sublattices. Note that the inverse of the Fourier transform is discrete. Making use of the Poisson summation rule $\sum_{R^{(\ell)}} e^{i k^{(l)} \cdot R^{(\ell)}} = A_{\text{BZ},\ell} \sum_{G^{(\ell)}} \delta_{k^{(l)}, G^{(\ell)}}$ and using the two-center approximation, we can rewrite the tight-binding Hamiltonian in momentum space:

$$H^l(k^{(l)}) = \begin{bmatrix} 0 & -t f_\ell(k^{(l)}) \\ t f_\ell^*(k^{(l)}) & 0 \end{bmatrix}, \tag{S7}$$

where $f_\ell(k^{(l)}) = \sum_i e^{i k^{(\ell)} \cdot s_i^{(\ell)}}$, and $s_i^{(\ell)}$ is the nearest-neighbor separation between A and B sites in the $\ell$-th layer, defined as $s_1^{(\ell)} = 1/3(a_1^{(\ell)} + a_2^{(\ell)}), s_2^{(\ell)} = 1/3(-2a_1^{(\ell)} + a_2^{(\ell)}), s_1^{(\ell)} = 1/3(a_1^{(\ell)} - 2a_2^{(\ell)})$, and $t$ is the tight-binding hopping parameter. The interlayer Hamiltonian can be written in momentum space as

$$V^{ij} = \sum_{\alpha,\beta} \int_{\text{BZ},i} dk^{(i)} \int_{\text{BZ},j} dk^{(j)} c_{i,k^{(i)},\alpha}^\dagger T_{ij}^{\alpha\beta}(k^{(i)}, k^{(j)}) c_{j,k^{(j)},\beta}, \tag{S8}$$

where

$$T_{ij}^{\alpha\beta}(k^{(i)}, k^{(j)}) = \frac{1}{A_{\text{u.c.}}} \sum_{G^{(i)},G^{(j)}} e^{i G^{(i)} \cdot \tau_\alpha^{(i)}} \tilde{t}_{\alpha\beta}^{ij}(k^{(i)} + G^{(i)}) e^{-i G^{(j)} \cdot \tau_\beta^{(j)}}$$
$$\times \delta_{k^{(i)} - G^{(i)}, k^{(j)} - G^{(j)}}, \tag{S9}$$

where $A_{\text{u.c.}}$ is the monolayer unit cell area and $\tilde{t}_{\alpha\beta}^{ij}$ is the Fourier component of the tight-binding hopping parameter [22]. The equation above gives two sets of scattering selection rules: $k^{(i)} - G^{(i)} = k^{(i+1)} - G^{(i+1)}$. In the bilayer case, the scattering selection rule is $k^{(1)} - G^{(1)} = k^{(2)} - G^{(2)}$, which forms a periodic lattice in momentum space with the periodicity of the bilayer moiré reciprocal lattice vector. However, with the additional layer, the two scattering selection rules give the following scattering selection rule:

$$\boldsymbol{k}^{(1)} - \boldsymbol{G}^{(1)} + \boldsymbol{G}^{(2)} = \boldsymbol{k}^{(3)} - \boldsymbol{G}^{(3)} + \boldsymbol{G'}^{(2)}, \tag{S10}$$

where $\boldsymbol{G}^{(2)}$ and $\boldsymbol{G'}^{(2)}$ are generally different. This selection connects additional momentum degrees of freedom that are uncoupled in the tBLG model.

The expressions that we have obtained above are exact. Further simplification can be made by taking the low-energy limit and expanding around $\boldsymbol{k}^{(\ell)} = \boldsymbol{q}^{(\ell)} + K_{L_\ell}$. In this limit, the intralayer term can be approximated by rotated linear Dirac Hamiltonians

$$H^1_{D(\boldsymbol{q}^{(1)})} = v_F \begin{bmatrix} 0 & e^{i\theta_{12}} q^{(1)}_+ \\ e^{-i\theta_{12}} q^{(1)}_- & 0 \end{bmatrix},$$

$$H^2_{D(\boldsymbol{q}^{(1)})} = v_F \begin{bmatrix} 0 & q^{(1)}_+ \\ q^{(1)}_- & 0 \end{bmatrix},$$

$$H^3_{D(\boldsymbol{q}^{(1)})} = v_F \begin{bmatrix} 0 & e^{-i\theta_{23}} q^{(1)}_+ \\ e^{i\theta_{23}} q^{(1)}_- & 0 \end{bmatrix}, \tag{S11}$$

where $q^{(l)}_\pm = q^{(l)}_x \pm q^{(l)}_y$, and $v_F$ is the Fermi velocity of monolayer graphene which we take to be $v_F = 0.8 \times 10^6$ m/s from the DFT calculated value. For the interlayer coupling, we keep the nearest neighbor coupling in momentum space:

$$T^{ij}_{\alpha\beta}(\boldsymbol{q}^{(i)}, \boldsymbol{q}^{(j)}) = \sum_{n=1}^{3} T^{q^{ij}_n}_{\alpha\beta} \delta_{\boldsymbol{q}^{(i)} - \boldsymbol{q}^{(j)}, -\boldsymbol{q}^{ij}_n}, \tag{S12}$$

where $\boldsymbol{q}^{ij}_1 = K_{L_i} - K_{L_j}$, $\boldsymbol{q}^{ij}_2 = \mathcal{R}^{-1}\left(\frac{2\pi}{3}\right) \boldsymbol{q}^{ij}_1$, $\boldsymbol{q}^{ij}_3 = \mathcal{R}\left(\frac{2\pi}{3}\right) \boldsymbol{q}^{ij}_1$ and $\mathcal{R}(\theta)$ is the counterclockwise rotation matrix by $\theta$. We take into account the out-of-plane relaxation by letting $t^{ij}_{AA} = t^{ij}_{BB} = \omega_0 = 0.07$ eV and $t^{ij}_{AB} = t^{ij}_{BA} = \omega_1 = 0.11$ eV due to the strengthened interaction between AB/BA sites from relaxation

$$T^{q^{ij}_1} = \begin{bmatrix} \omega_0 & \omega_1 \\ \omega_1 & \omega_0 \end{bmatrix}, T^{q^{ij}_2} = \begin{bmatrix} \omega_0 & \omega_1 \bar{\phi} \\ \omega_1 \phi & \omega_0 \end{bmatrix}, T^{q^{ij}_3} = \begin{bmatrix} \omega_0 & \omega_1 \phi \\ \omega_1 \bar{\phi} & \omega_0 \end{bmatrix}, \tag{S13}$$

Where $\phi = e^{\frac{i2\pi}{3}}$ and $\bar{\phi} = e^{-\frac{i2\pi}{3}}$. The Hamiltonian in momentum space can then be formally written as

$$H_K(\boldsymbol{k}) = \begin{bmatrix} H^1_D(\boldsymbol{k}) & T^{12} & 0 \\ T^{12\dagger} & H^2_D(\boldsymbol{k}) & T^{23} \\ 0 & T^{23\dagger} & H^2_D(\boldsymbol{k}) \end{bmatrix}. \tag{S14}$$

We compute the density of states (DOS) by summing up the states at each energy in the bilayer moiré Brillouin zone of layers 1 and 2 in a uniform grid. We approximate the delta function by a Gaussian function:

$$g(E) = \sum_{k \in BZ, 12} \delta(E - E_k) \approx N \sum_{\alpha=A,B} \sum_{\ell=1,2} \sum_n \int_{\Gamma^{\ell,\ell+1}} |\psi_{n,k}|^2 e^{-\frac{(E-E_k)^2}{2\sigma^2}} d^2k, \tag{S15}$$

where $\Gamma^{\ell,\ell+1}$ is the bilayer moiré cell between layers $\ell$ and $\ell+1$, $\psi_{n,k}$ is the wavefunction with the band index $n$ and momentum $k$, $\sigma$ is the full width half maximum of the Gaussian, and $N$ is a normalization constant.

In order to compare the magnitude of the van Hove peaks between tTLG and tBLG, we need to properly normalize the DOS. For a given cutoff radius, we first calculate the DOS of the intralayer Hamiltonian only, which reduces to three independent copies of monolayer graphene. Near the charge-neutrality point (CNP), the DOS per nm$^2$ is given by

$$g(E) = \frac{6}{\pi v_F^2}|E|, \qquad (S16)$$

where the prefactor includes a factor of 3 from the number of layers as well as a factor of 4 from spin and valley degeneracies. We then obtain a normalization constant $N$ by fixing the prefactor to the expected slope given in Eq. (S16) and use the same constant for the DOS of the full Hamiltonian, including both the intralayer and interlayer terms.

All the presented DOS in tTLG were computed using a grid size of 32 by 32. This choice is justified with higher grid samplings, up to 162 by 162 (26, 244 momenta), that gave similar results.

## S9. Relaxed Total Energy Landscape Calculation

To calculate the relaxed total energy landscape, we employ a continuum relaxation model in local configuration space [23]. Configuration space describes the local environment of every position in layer $L_\ell$ and bypasses a periodic approximation [24]. Every position in real space in $L_\ell$ can be uniquely parametrized by three shift vectors $\boldsymbol{b}^{i \to j}$ for $j = 1, 2, 3$ that describe the relative position between a given point $\boldsymbol{r}$ with respect to all three layers. Note that $\boldsymbol{b}^{i \to j} = \boldsymbol{0}$ if $i = j$ since the separation between a position with itself is 0, which leads to a four-dimensional configuration space.

For a given real space position $\boldsymbol{r}$, the following linear transformation relates $\boldsymbol{r}$ and $\boldsymbol{b}^{i \to j}$ in layer $i$ with respect to layer $j$, and the following linear transformation maps the relaxation from the local configuration to the real space position $\boldsymbol{r}$:

$$\boldsymbol{b}^{i \to j}(\boldsymbol{r}) = (E_j^{-1} E_i - 1)\boldsymbol{r}, \qquad (S17)$$

where $E_i$ and $E_j$ are the unit cell vectors of layers $i$ and $j$ respectively, rotated by $\theta_{ij}$. In the trilayer system, there is no simple linear transformation between real and configuration space. The relation between the displacement field defined in real space, $\boldsymbol{U}^{(i)}(\boldsymbol{r})$, and in configuration space, $\boldsymbol{u}^{(i)}(\boldsymbol{b})$, can be found by evaluating $\boldsymbol{u}^{(j)}(\boldsymbol{b})$ at the corresponding $\boldsymbol{b}^{i \to j}(\boldsymbol{r})$ and $\boldsymbol{b}^{i \to j}(\boldsymbol{r})$ with Eq. (6) to obtain $\boldsymbol{U}^{(i)}(\boldsymbol{r}) = \boldsymbol{u}^{(i)}\left(\boldsymbol{b}^{(i \to j)}(\boldsymbol{r}), \boldsymbol{b}^{(i \to k)}(\boldsymbol{r})\right)$, where $j, k \neq i$ and $j < k$.

The relaxed energy has two contributions, intralayer and interlayer energies:

$$E^{\text{tot}}(\boldsymbol{u}^{(1)}, \boldsymbol{u}^{(2)}, \boldsymbol{u}^{(3)}) = E^{\text{intra}}(\boldsymbol{u}^{(1)}, \boldsymbol{u}^{(2)}, \boldsymbol{u}^{(3)}) + E^{\text{inter}}(\boldsymbol{u}^{(1)}, \boldsymbol{u}^{(2)}, \boldsymbol{u}^{(3)}), \qquad (S18)$$

where u^(l) is the relaxation displacement vector in layer l. To obtain the relaxation pattern, we minimize the total energy with respect to the relaxation displacement vector.

We model the intralayer coupling based on linear elasticity theory:

$$E^{\text{intra}}(\boldsymbol{u}^{(1)}, \boldsymbol{u}^{(2)}, \boldsymbol{u}^{(3)}) = \sum_{l=1}^{3} \frac{1}{2}[\left(\partial_x u_x^{(l)} + \partial_y u_y^{(l)}\right)^2$$

$$+ K\left(\left(\partial_x u_x^{(l)} - \partial_y u_y^{(l)}\right)^2 + \left(\partial_x u_y^{(l)} + \partial_y u_x^{(l)}\right)^2\right) d\boldsymbol{b}, \tag{S19}$$

where G and K are shear and bulk moduli of monolayer graphene, which we take to be G = 47352 meV/unit cell, K = 69518 meV/unit cell [23,24] Note that the gradient in Eq. (S19) is with respect to the real space position $\boldsymbol{r}$.

The interlayer energy accounts for the energy cost of the layer misfit, which is described by generalized stacking fault energy (GSFE) [25,26], obtained using first principles Density Functional Theory (DFT) with the Vienna Ab initio Simulation Package (VASP) [27–29]. GSFE is the ground state energy as a function of the local stacking with respect to the lowest energy stacking between a bilayer. For bilayer graphene, GSFE is maximized at the AA stacking and minimized at the AB stacking. Letting $\boldsymbol{b} = (b_x, b_y)$ be the relative stacking between two layers, we define the following vector $\boldsymbol{v} = (v, w) \in [0, 2\pi]^2$:

$$\begin{pmatrix} v \\ w \end{pmatrix} = \frac{2\pi}{a_0} \begin{bmatrix} \sqrt{3}/2 & -1/2 \\ \sqrt{3}/2 & 1/2 \end{bmatrix} \begin{pmatrix} b_x \\ b_y \end{pmatrix}, \tag{S20}$$

where $a_0 = 2.4595$ Å is the graphene lattice constant. We parametrize the GSFE as follows:

$$\begin{aligned} V_{(j\pm)}^{GSFE} &= c_0 + c_1(\cos(v) + \cos(w) + \cos(v+w)) \\ &+ c_2(\cos(v+2w) + \cos(v-w) + \cos(2v+w)) \\ &+ c_3(\cos(2v) + \cos(2w) + \cos(2v+2w)), \end{aligned} \tag{S21}$$

where we take $c_0 = 6.832$ meV/cell, $c_1 = 4.064$ meV/cell, $c_2 = -0.374$ meV/cell, $c_3 = -0.0095$ meV/cell [23]. The van der Waals force is implemented through the vdW-DFT method using the SCAN+rVV10 functional [29]. Note that we amplify the GSFE by a factor of 10. Physically, amplifying the GSFE enhances the strength of relaxation. It has been shown that the energy difference between AA and AB stackings can vary by a factor of 4 depending on the van der Waals functionals used [25]. In terms of $V_{(l\pm)}^{GSFE}$, the total interlayer energy can be expressed as follows:

$$E^{\text{inter}} = \frac{1}{2} \int V_{1+}^{\text{GSFE}}(\boldsymbol{B}^{1\to 2}) \, d\boldsymbol{b}$$

$$+ \frac{1}{2} \int \left[V_{2-}^{\text{GSFE}}(\boldsymbol{B}^{2\to 1}) + V_{2+}^{\text{GSFE}}(\boldsymbol{B}^{2\to 3})\right] d\boldsymbol{b} + \frac{1}{2} \int V_{3-}^{\text{GSFE}}(\boldsymbol{B}^{3\to 2}) \, d\boldsymbol{b}, \tag{S22}$$

where $\boldsymbol{B}^{(i\to j)} = \boldsymbol{b}^{(i\to j)} + \boldsymbol{u}^{((j))} - \boldsymbol{u}^{((i))}$ is the relaxation modified local shift vector. Note that we neglect the interlayer coupling between layers 1 and 3. The total energy is obtained by summing over uniformly sampled configuration space. In this work, we discretize the four-dimensional configuration space by $54 \times 54 \times 54 \times 54$.